\documentclass[conference]{IEEEtran}
\IEEEoverridecommandlockouts
\usepackage{cite}
\usepackage{amsmath,amssymb,amsfonts}
\usepackage{booktabs}
\usepackage{multirow}
\usepackage{algorithmic}
\usepackage{graphicx}
\usepackage{textcomp}
\usepackage{xcolor}
\usepackage{hyperref}
\usepackage{url}
\def\BibTeX{{\rm B\kern-.05em{\sc i\kern-.025em b}\kern-.08em
    T\kern-.1667em\lower.7ex\hbox{E}\kern-.125emX}}
\begin{document}

\newcommand{\ssn}{\(\,^{\mbox{-}}\)} 
\newcommand{\ssa}{\ensuremath{{}^{\textcolor{black}{\circ}}}} 
\newcommand{\ssb}{\ensuremath{{}^{\textcolor{black}{\bullet}}}} 

\title{Leveraging Large Language Models for Medical Information Extraction and Query Generation}

\author{\IEEEauthorblockN{Georgios Peikos}
\textit{University of Milano-Bicocca}\\
Milan, Italy \\
georgios.peikos@unimib.it
\and
\IEEEauthorblockN{Pranav Kasela}
\textit{University of Milano-Bicocca}\\
Milan, Italy \\
pranav.kasela@unimib.it
\and
\IEEEauthorblockN{Gabriella Pasi}
\textit{University of Milano-Bicocca}\\
Milan, Italy \\
gabriella.pasi@unimib.it}

\maketitle

\begin{abstract}
This paper introduces a system that integrates large language models (LLMs) into the clinical trial retrieval process, enhancing the effectiveness of matching patients with eligible trials while maintaining information privacy and allowing expert oversight. We evaluate six LLMs for query generation, focusing on open-source and relatively small models that require minimal computational resources. Our evaluation includes two closed-source and four open-source models, with one specifically trained in the medical field and five general-purpose models. We compare the retrieval effectiveness achieved by LLM-generated queries against those created by medical experts and state-of-the-art methods from the literature. Our findings indicate that the evaluated models reach retrieval effectiveness on par with or greater than expert-created queries. The LLMs consistently outperform standard baselines and other approaches in the literature. The best performing LLMs exhibit fast response times, ranging from 1.7 to 8 seconds, and generate a manageable number of query terms (15-63 on average), making them suitable for practical implementation. Our overall findings suggest that leveraging small, open-source LLMs for clinical trials retrieval can balance performance, computational efficiency, and real-world applicability in medical settings.
\end{abstract}

\section{Introduction}
\label{sec:intro}
Hospitals and other medical institutions store patient-related information in electronic health records (EHRs), which include clinical notes, discharge summaries, laboratory test results, and other valuable information.
Beyond their primary medical applications, these records serve multiple secondary purposes aiming to enhance healthcare services~\cite{DBLP:journals/csur/SarwarSCZAHVC23}.
One significant secondary use is their role in identifying suitable clinical trials for patient participation, thus advancing medical research and offering patients potential new treatment options.

Clinical trials are the established scientific method for evaluating new biological agents, drugs, devices, or procedures aimed at preventing or treating diseases in human populations~\cite{fleming1996surrogate}.
A significant challenge in clinical trials is recruiting sufficient participants, which can delay trials, lead to failures, and limit their generalization~\cite{gul2010clinical}.
The enrollment process begins when healthcare providers construct search queries based on patient information usually mentioned in clinical notes or discharge summaries~\cite{jain2019conceptual}.
However, clinical summaries often exhibit lengthiness, lack of structure, and various textual complexities, including medical jargon, abbreviations, and occasional grammatical errors.
These characteristics increase the effort required from healthcare providers, making identifying suitable clinical trials more time-consuming and complex.

To manage these complexities, several approaches in the literature utilize rule-based, hybrid, or neural-based Natural Language Processing (NLP) methods~\cite{DBLP:journals/kais/LandolsiHR23,DBLP:journals/ijmi/HobensackSSBT23}.
These methods are designed to efficiently extract valuable information from clinical summaries, thereby reducing human effort across a multitude of tasks, including clinical trials retrieval~\cite{DBLP:conf/ecir/PeikosAPV23}.
Recent studies show that leveraging large language models (LLMs), such as GPT3.5 and GPT4, significantly enhances patient enrollment by improving clinical trials retrieval effectiveness~\cite{DBLP:journals/jbi/ParkFTZICFSGSW24,peikos2023utilizing}.
However, the practical deployment of these models in real-world settings presents challenges due to their resource-intensive computational demands and privacy risks associated with handling sensitive medical data.
Consequently, their application in medical institutions, particularly those with limited resources, is hampered~\cite{zuhair2024exploring}.
Furthermore, given that the previous studies focus on GPT-based models, the efficacy of small, open-source models, which are more feasible for real-world scenarios, remains to be investigated.

In this paper, we propose a system that integrates LLMs into the clinical trial retrieval process, enhancing the  effectiveness of matching patients with eligible clinical trials while maintaining information privacy and allowing expert oversight.
We assess six LLMs, mainly small and open-source, on their ability to produce patient-tailored queries that enhance the retrieval effectiveness of eligible clinical trials when used in searches.
We compare the retrieval effectiveness achieved by LLM-generated queries with those created by medical experts, large closed-source LLMs, and state-of-the-art methods proposed in the literature.

The paper is organized as follows. Section~\ref{sec:lit_review} provides an overview of studies related to medical information extraction (IE) and clinical trials retrieval.
Section~\ref{sec:methodology} details our approach to leverage LLMs for extracting patient-specific information and generating queries for clinical trials retrieval.
Section~\ref{sec:exp_setup} outlines the experimental setup and Section~\ref{sec:results} presents the results of our experiments.
Finally, in Section~\ref{sec:disc}, we discuss the primary outcomes of our study, while Section~\ref{sec:conc_fd} concludes our work.
\section{BACKGROUND AND RELATED WORKS}
\label{sec:lit_review}
LLMs, mainly closed-source models like GPT, have been extensively utilized in the medical domain in numerous applications such as generating discharge summaries~\cite{patel2023chatgpt} and medical notes~\cite{DBLP:journals/jms/CascellaMBB23}, triaging patients~\cite{sezgin2022operationalizing}, and anonymizing medical texts~\cite{DBLP:journals/corr/abs-2303-11032}.
While the results from these studies are encouraging, LLM limitations (e.g. hallucinations, lack of general reasoning capability) raise concerns regarding their adoption in medical applications~\cite{harrer2023attention,patel2023chatgpt}.
Therefore, as emphasized in~\cite{yang2023large}, it is essential to ensure that AI-generated content undergoes thorough review to mitigate potential risks and maintain the high standards of patient care and medical ethics.
\subsection{Information Extraction from Clinical Notes}
Information Extraction systems in clinical settings have 
been extended
from rule-based approaches to advanced machine learning techniques, including deep neural networks and transformer architectures like ClinicalBERT~\cite{DBLP:journals/jamia/WuRDDJSSWWXZX20, DBLP:journals/jbi/KreimeyerFPAHJF17, DBLP:journals/jbi/WangWRMSALZMSL18, alsentzer-etal-2019-publicly, beltagy2019scibert,DBLP:journals/kais/LandolsiHR23,hahn2020medical}.
State-of-the-art performance in various clinical NLP tasks, such as named entity recognition and information extraction from medical texts, has been achieved by LLMs, mainly the GPT3.5 model~\cite{hu2023zero,DBLP:conf/emnlp/AgrawalHLKS22}.
Specifically, for information extraction from lung cancer and pediatric osteosarcoma pathology reports, ChatGPT achieves 89\% accuracy in extracting structured data, with most errors occurring due to the model's missing specialized terminology~\cite{huang2024critical}.
Similarly, ChatGPT has been employed to extract and standardize patient information from breast cancer pathology and ultrasound reports, with findings highlighting the time and cost-effectiveness of using LLMs for this task~\cite{choi2023developing}.
Given the numerous studies leveraging LLMs to process EHRs, particularly unstructured clinical texts, we refer interested readers to the comprehensive scoping review presented in~\cite{li2024scoping}.
The review analyzes 329 studies on applying LLMs in EHRs, covering trends, data resources, model types, and various NLP tasks.
\subsection{Clinical Trials Retrieval}
The task of clinical trials retrieval, which aims to enhance patient enrollment, is formulated as follows.
Given a patient's admission note containing relevant medical information (the query), the system retrieves from a collection of clinical trials (the document collection) those trials for which the patient is eligible to participate~\cite{soboroff2021overview,roberts2022overview}.
Researchers have proposed various retrieval approaches to address this task. These approaches combine standard retrieval methods with multiple query variations~\cite{DBLP:conf/sigir/KoopmanZ16}, explore query expansion and reduction techniques~\cite{DBLP:conf/sigir/AgostiN019}, and utilize neural models for document re-ranking~\cite{DBLP:journals/jbi/RybinskiXK20}. 
Particular research attention has been given to this task through the TREC initiative, specifically in the TREC Clinical Trials 2021 (TREC21) and 2022 (TREC22) tracks \cite{soboroff2021overview,roberts2022overview}, which have enabled researchers to evaluate and compare different approaches.

The state-of-the-art (SoA) approach from TREC21 employs a multi-stage retrieval process~\cite{10.1145/3477495.3531853}.
Initially, Neural Query Synthesis (NQS), which is a zero-shot document expansion method, generates forty sentence-length queries from a given clinical note.
These generated queries, along with the original note, serve as inputs for forty-one separate retrieval pipelines that leverage the BM25~\cite{DBLP:conf/trec/RobertsonWJHG94} and RM3~\cite{DBLP:conf/trec/JaleelACDLLSW04} retrieval models.
The resulting document rankings from these pipelines are then fused into a single ranking.
In the second retrieval phase, a neural re-ranker based on a Mono-T5 model, specifically fine-tuned for clinical trial retrieval, is applied to produce the 
final ranking.
The TREC22 SoA approach, developed by the same research team, also utilizes the Mono-T5 model~\cite{roberts2022overview}.
However, due to limited published experimental details, a comprehensive analysis of this approach is not feasible.
Other approaches published within the TREC21 and TREC22 tracks utilize various techniques for query generation and retrieval.
These methods employ unsupervised information extraction and query expansion techniques, such as KeyBERT~\cite{grootendorst2020keybert}, leverage medical knowledge bases like UMLS~\cite{bodenreider2004unified}, or use syntactical rules to generate queries for trials retrieval~\cite{peikos2022unimib,kusa2021dossier}.

Extracting key patient information to generate search queries by combining rule-based methods with domain-specific pre-trained transformer models has been shown to enhance clinical trials retrieval.
Nonetheless, only specific query formulations, particularly those that exclude patient data related to family medical history, tend to produce better retrieval results~\cite{DBLP:conf/ecir/PeikosAPV23}.
Recent research incorporating ChatGPT has yielded even greater retrieval performance.
The evidence suggest that patient information extracted using ChatGPT can significantly increase retrieval effectiveness in the context of the studied search task, exceeding the performance of the SoA in TREC21 and TREC22, and, in some instances, surpassing the quality of queries crafted by medical experts~\cite{peikos2023utilizing}.

The system we propose in this study addresses essential concerns in the medical AI field.
Specifically, in line with recommendations from previous studies~\cite{harrer2023attention,patel2023chatgpt,yang2023large}, the system allows medical experts to thoroughly examine AI-generated patient-related queries, thereby mitigating potential risks associated with automated systems in healthcare.
Moreover, we experiment with small open-source language models that can be run locally within medical institutions, significantly enhancing data security—a critical consideration in handling sensitive patient information.
\section{Proposed System Architecture}
\label{sec:methodology}
\begin{figure}[!t]
    \centering
    \includegraphics[width=.98\linewidth]{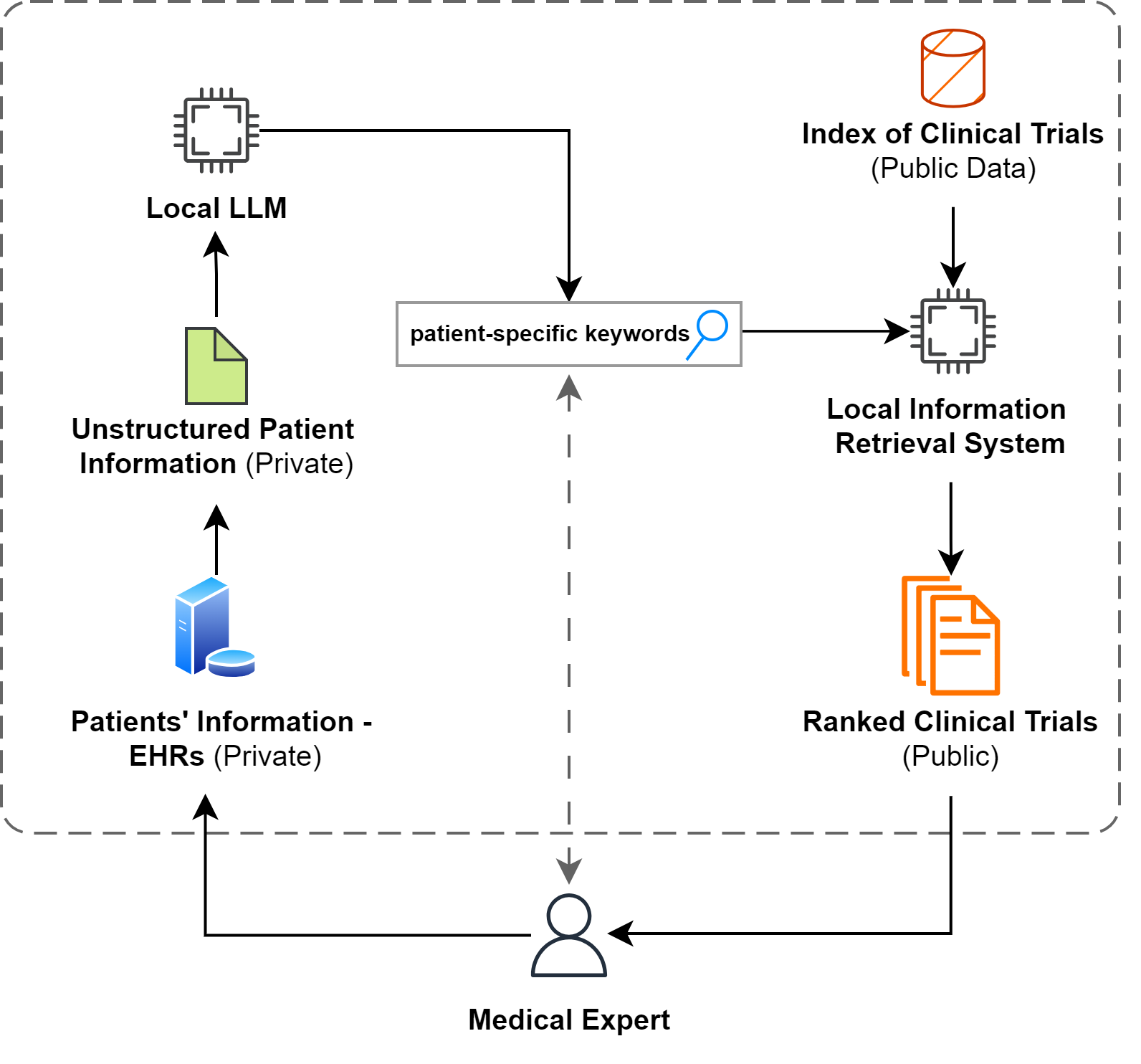}
    \caption{Real-world application of the proposed system.}
    \label{fig:method_overview}
\end{figure}

Integrating LLMs into the clinical trial retrieval process necessitates alterations to the standard process presently utilized in medical institutions~\cite{jain2019conceptual, DBLP:journals/kais/LandolsiHR23}.
Figure~\ref{fig:method_overview} depicts this modified procedure, which leverages LLM's capabilities while maintaining patient privacy and incorporating expert oversight.

The process consists of the following steps.
Initially, a medical expert selects a patient of interest from the institution's secured record system, and the patient's information, in form of a clinical summary, are selected for further processing.
The unstructured clinical note is incorporated into a prompt, such as the one presented in Figure~
\ref{fig:prompt}, and constitutes the input to a locally running LLM.
The model operates within the hospital's secure environment so it eliminates transmissions of private information to third parties, ensuring data privacy.
The LLM processes the clinical note and generates a query following the prompt instructions.
In a real-world application of the system, the medical expert can accept the generated query as-is or modify it with further domain-specific knowledge.
This interaction, illustrated with dashed arrows in Figure~\ref{fig:method_overview}, allows for human oversight and refinement of the AI-generated content, enhancing human-AI collaboration~\cite{reverberi2022experimental}.
Once finalized, the query is input into an Information Retrieval (IR) system.
This system leverages an index of publicly available clinical trials, and using an IR model ranks these trials based on their 
relevance
to the provided query.
The process concludes with the medical expert being presented a ranked list of clinical trials, with the most relevant trial appearing first in the results.
Ultimately, the expert reviews the retrieved clinical trials and selects the most appropriate for the patient of interest.
\section{Experimental Setup}
\label{sec:exp_setup}
This section details the main components of our methodology, including the representation of patient information, the selected LLMs and their configurations, the processing of the generated queries, and the implementation of the IR system for clinical trials retrieval.
\subsection{Stored Patient Information}
The storage and transmission of patient information, whether in cloud services or hospital databases, presents significant challenges regarding data security, privacy protection, and regulatory compliance, e.g., the Health Insurance Portability and Accountability Act (HIPAA)\footnote{\href{https://www.cdc.gov/phlp/publications/topic/hipaa.html}{Health Insurance Portability and Accountability Act of 1996 (HIPAA).}}~\cite{shah2020secondary}.
In the empirical evaluation we conduct to assess our proposed methodology, we utilize synthetic patient cases, i.e. unstructured patient clinical notes, created by individuals with medical training, rather than real patient information.
Each note contains various patient-related information, including key details such as patient demographics, medical history, current diagnoses, medications, and relevant laboratory results (e.g., underlying medical problems and family history).
An example note is depicted in Figure~\ref{fig:prompt}.
\subsection{Large Language Models}
In our experiments, we prioritize the use of 
small, open-source LLMs that can run locally, addressing their practical implications when deployed in medical institutions, especially those found in developing nations~\cite{zuhair2024exploring}.
Our approach aims to overcome resource constraints and infrastructure limitations often encountered in these settings.
Specifically, our experiments utilized a maximum of 30GB of memory running on a machine equipped with an A6000 GPU, which has a total capacity of 48GB.
This setup can be used in hospitals, allowing them to derive significant benefits regarding infrastructure requirements, operational efficiency, and information privacy.
We evaluate four open-source LLMs whose parameters range from 4 billion to 14 billion and can be deployed on a single GPU.
Specifically, we use:
\begin{itemize}
    \item Qwen2-7B-Instruct.
    Released in June 2024 by the Qwen Team, this model has been trained on 29 languages, including English and Chinese. It supports a context length of up to 128k tokens and has been trained on a dataset of approximately 7 trillion tokens, collected by the Qwen Team~\cite{yang2024qwen2technicalreport}.
    \item Phi3-medium-4k-Instruct\footnote{\url{https://huggingface.co/microsoft/Phi-3-medium-4k-instruct}}. 
    Released by Microsoft in May 2024, this lightweight, state-of-the-art model has been trained on datasets comprising both synthetic and meticulously filtered publicly available website data, selected for their high-quality and dense reasoning properties. The dataset used totals 4.8 trillion tokens, and the training process spanned 42 days. The medium version of the model has 14 billion parameters~\cite{abdin2024phi3technicalreport}.
    \item Phi-3-mini-4k-Instruct\footnote{\url{https://huggingface.co/microsoft/Phi-3-mini-4k-instruct}}.
    This model is a scaled-down variant of Phi3-medium-4k-Instruct, featuring 3.8 billion parameters. It underwent identical training using the same procedures and data as its medium-sized counterpart, but with a shorter training period of only 10 days.
    \item Medical-Llama3-8B\footnote{\url{https://huggingface.co/ruslanmv/Medical-Llama3-8B}}.
    This model is a fine-tuned version of the Llama3 8B model, trained to provide answers to medical queries.
    Its training leverages the extensive knowledge embedded within the AI Medical Chatbot dataset\footnote{\url{https://huggingface.co/datasets/ruslanmv/ai-medical-chatbot}}.
\end{itemize}

To maintain comparability with existing literature, which leverages closed-source and larger LLMs, we have included GPT3.5 and GPT4 in our evaluation.

In our study, we prioritized the use of the instruct model versions.
While non-instruct versions of LLMs are trained on diverse data and can generate responses based on general language patterns, they may lack precision in following detailed instructions.
Instruct model versions undergo additional fine-tuning processes, including supervised learning and direct preference optimization, specifically designed to enhance their ability to follow instructions accurately.
This post-training optimization makes them particularly suitable for specialized tasks.
Furthermore, we leveraged the domain-specific model trained on medical literature and data due to its ability to generate outputs that are highly contextually relevant to the medical domain.
\begin{figure}[!t]
    \centering
    \includegraphics[width=\linewidth]{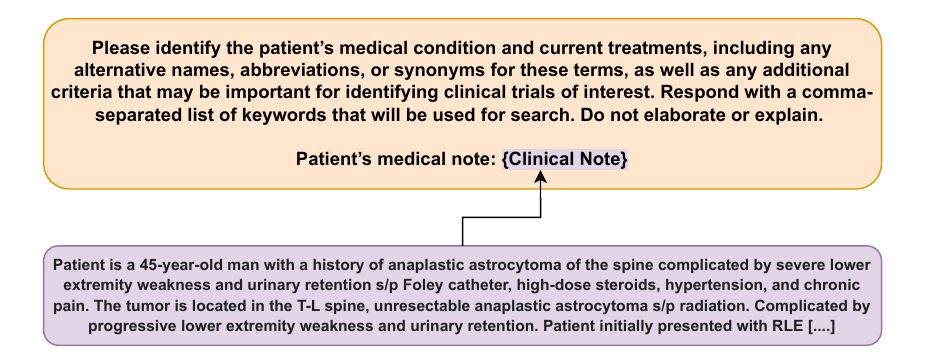}
    \caption{Example of a clinical trial (in bottom) with the prompt utilized for the query generation (on top).}
    \label{fig:prompt}
\end{figure}

The prompt used to input the patient's clinical note into the LLMs is depicted in Figure~\ref{fig:prompt}.
This prompt builds upon the best performing configuration for this task, as identified in~\cite{peikos2023utilizing}.
\subsection{Generated Query Processing}
The generated queries undergo processing to mitigate common errors. We remove numbered list patterns often produced by LLMs while preserving non-list numerical values (e.g. age, dosages). Punctuation and stopwords are eliminated, and Porter stemming is applied. These steps prepare the query to be input into the IR system.

Our empirical evaluation is conducted without real domain experts involved.
Therefore, we compared two approaches for query processing:
the first one utilizes only the generated query, as it is, as input to the IR system, and the second one that employs a concatenation of the original query with the generated query.
This experimental design enabled us to assess the retrieval effectiveness of using the generated query alone versus combining it with the original query, indicating whether the generated query is sufficient by itself, or if retaining the original query provides additional value in the retrieval process.
\subsection{Information Retrieval System}
The local IR system requires an index to efficiently search and retrieve clinical trial documents.
These documents are publicly available, for example, on ClinicalTrials.gov\footnote{\href{https://clinicaltrials.gov/}{ClinicalTrials.gov, accessed on 03/07/2024.}}, which is a mandatory registry for clinical trials in the United States. 
A clinical trial document follows a structured format that includes sections such as title, studied condition, summary, detailed description, and eligibility criteria, among others.
While all these sections are essential during retrieval, the eligibility section is particularly crucial as it contains information (i.e., inclusion/exclusion criteria) that may determine whether a given patient is eligible to participate in the trial. 

To index clinical trials, we employed PyTerrier~\cite{pyterrier2020ictir} with its default parameters, including Porter stemming and stopword removal.
We indexed the title, studied condition, summary, detailed description, and eligibility criteria, as well as the sections of each clinical trial document.
To retrieve eligible clinical trials based on patient-specific generated queries, we employ the widely used BM25 model~\cite{DBLP:conf/trec/RobertsonWJHG94}.
Specifically, we used PyTerrier's implementation of BM25 with its default parameters across all retrieval experiments conducted in this work.
This approach ranks clinical trials with respect to their topical similarity to the patient-specific query, producing a ranking in which the first trial is the most relevant.
\section{EVALUATION}
\label{sec:results}
This section presents the results of our experimental evaluation, detailing the retrieval effectiveness of queries generated by LLMs compared to those created by medical experts.
Additionally, it showcases the ability of open-source small LLMs to perform this task compared to larger models like GPT4 and the SoA approaches for this task.
\subsection{Benchmark Collections and Evaluation Measures}
The comparison with expert-created queries utilizes the benchmark collection introduced by Koopman and Zuccon~\cite{DBLP:conf/sigir/KoopmanZ16}.
The collection comprises sixty queries simulating patient cases in the form of admission summaries. For each summary, a set of keywords created by four medical experts is provided.
These keywords represent the terms the experts would have used to search for relevant clinical trials for the patients described in the summaries.
The document collection consists of clinical trials extracted from ClinicalTrials.gov.

The evaluation of open-source small LLMs 
for
query generation 
in the considered
task utilizes the two benchmark collections introduced in TREC21~\cite{soboroff2021overview} and TREC22~\cite{roberts2022overview}.
Similar to the previously mentioned one, these collections consist of patient case descriptions (queries) and clinical trials from ClinicalTrials.gov (documents).

The relevance assessment methodology is identical for all three collections. For each clinical note, a trial is categorized as \textit{eligible}, \textit{excludes}, or \textit{not relevant}. An \textit{eligible} label indicates that the patient can participate in the trial. \textit{Excludes} means that the patient 
fulfills the conditions targeted by the trial, 
but the exclusion criteria make the patient ineligible. \textit{Not relevant} signifies that the patient does not meet the criteria for the trial.

Following the official evaluation guideline for these tasks, we report the nDCG@10, P@10 (Precision at 10), mean reciprocal rank (MRR), Bpref, and R@25 (recall at 25).
For a detailed description of the employed evaluation measures, please refer to~\cite{DBLP:books/daglib/0021593}
and~\cite{DBLP:journals/ftir/MitraC18}.
Also, besides the nDCG measure, the remaining measures assume only \textit{eligible} trials as relevant, i.e. trials assessed as \textit{excludes} are also considered \textit{not relevant}.
In the benchmark collection introduced in~\cite{DBLP:conf/sigir/KoopmanZ16}, the number of assessed documents is limited, and therefore, the evaluation might be less reliable for new systems, as mentioned by the authors.
To overcome this issue, we evaluate the retrieval performance by employing the condensed measures approach proposed in~\cite{DBLP:conf/sigir/Sakai07} that omits the unjudged documents from the evaluation.

\subsection{Comparing LLM-generated to Expert-created Queries}
We evaluate the performance of the generated queries by comparing them with those produced by domain experts. Our analysis follows the methodology outlined in a previous study~\cite{peikos2023utilizing}, with an expansion to include closed-source models.

The obtained results are presented in Table~\ref{tab:llm_vs_human}, where \textit{Expert A}, \textit{B}, \textit{C}, \textit{D} refer to the experiments that leverage the keyword-based queries created by each expert.
\textit{Expert Combined} refers to an experiment that concatenates all unique keywords created by all experts, removing any duplicates.
We assume that the \textit{Expert Combined} query accumulates the knowledge of different experts in a single representation.
The remaining experiments leverage the LLM-generated queries. To retrieve clinical trials, we use the BM25 model for all the reported experiments.

The retrieval performance achieved by the expert-created queries shows high variability, while the \textit{Expert Combined} query representation does not 
outperform the best individual expert, suggesting that a simple combination of expert knowledge may not be optimal.
Open-source models, particularly Qwen2-7B-Instruct and Phi3-medium-4k-Instruct, achieve a retrieval performance comparable to that of both larger closed-source models like GPT3.5 and several experts.
While these models do not consistently outperform the best expert-created queries, their retrieval effectiveness is on par, and it surpasses that achieved by the \textit{Expert Combined} and some individual experts.
The domain-specific Medical-Llama3-8B achieves the lowest performance among LLMs in our evaluation.
We conducted a qualitative analysis of the generated queries to investigate the reasons behind this performance and understand why other models perform better.
Figure~\ref{fig:query_comparison} presents representative examples of LLM-generated queries, mainly illustrating the challenges faced by the Medical Llama3 model in following prompt instructions and its difficulty in generating valuable keywords.
While the difficulty in adhering to instructions is somehow expected, given that it is not fine-tuned for instruction-following, its inability to generate valuable keywords is less 
foreseeable
as it is a domain-specific model.
\begin{table}[!t]
\centering
\caption{Retrieval performance achieved by LLM-generated and expert-created queries using BM25.}
\label{tab:llm_vs_human}
\begin{tabular}{c|c|cc}
\multicolumn{2}{c|}{Experiment} & Bpref & P@10 (Condensed) \\ \hline
\multirow{5}{*}{Experts} & \textit{Expert A} & .117 & .140 \\
 & \textit{Expert B} & .093 & .120 \\
 & \textit{Expert C} & .059 & .100 \\
 & \textit{Expert D} & .116 & .138 \\
 & \textit{Experts Combined} & .090 & .110 \\ \hline
\multirow{2}{*}{Closed Source} & GPT4 & .098 & .118 \\
 & GPT3.5 & .107 & .131 \\ \hline
\multirow{4}{*}{Open Source} & Qwen2-7B-Instruct & .104 & .130 \\
 & Phi3-medium-4k-Instruct & .102 & .134 \\
 & Phi3-mini-4k-Instruct & .086 & .110 \\
 & Medical-Llama3-8B & .067 & .081
\end{tabular}
\end{table}
Overall, open-source LLMs, despite their significantly smaller size, prove to be effective for clinical trials retrieval, successfully balancing performance and computational efficiency.
\subsection{Comparing LLM-generated Queries with SoA approaches}
Tables~\ref{tab:clinical_trial_trec21} and~\ref{tab:clinical_trial_trec22} 
report
the retrieval performance of the considered baseline, the evaluated LLMs, and the SoA approaches across the employed collections.

As a baseline (\textit{Original Query}), we use the original clinical note with query pre-processing before input into our BM25-based IR system.
The statistical significance of the difference for all reported measures against the baseline is tested using a paired t-test with Bonferroni correction for multiple comparisons, at a significance level of 0.05, denoted in the tables as (\ssa)~\cite{carterette2012multiple}.
We also evaluate the RM3 model (\textit{Original Query+RM3})~\cite{DBLP:conf/trec/JaleelACDLLSW04}, a standard approach for query expansion based on pseudo-relevance feedback.
RM3 is implemented using PyTerrier, with ten feedback documents and twenty expansion terms concatenated to the original query.
For the LLM-based experiments, as detailed in Section~\ref{sec:exp_setup}, we implemented two approaches: using only LLM-generated queries and concatenating original queries with LLM-generated ones.
The concatenation approach consistently 
outperforms the LLM-generated queries alone.
In the Tables \ref{tab:llm_vs_human}, \ref{tab:clinical_trial_trec21}, and \ref{tab:clinical_trial_trec22}, we report only the results of the concatenation method due to its higher performance values.
Additionally, the tables report the first-stage retrieval performance of the best performing approaches from the TREC 2021 and 2022 tracks to facilitate direct comparison with the employed LLM models.
Finally, we report the average processing time per query in seconds, and the average number of terms across all generated queries for each experiment.
It is important to note that closed-source models were accessed through their API and that we have not utilized flash attention for the phi3 models.
These implementation decisions may have impacted the response times of these models.
\begin{table*}[!t]
    \centering
    \caption{Results on the TREC21 benchmark using BM25 model. In boldface the best results overall. Blank cells indicate that the value can not be calculated or is not reported.}
    \label{tab:clinical_trial_trec21}
    \resizebox{\linewidth}{!}{%
    \begin{tabular}{cccccccp{7em}p{5em}}
         & Experiment & nDCG@10 & Bpref & P@10 & R@25 & MRR & Avg. processing time (seconds) & Average terms \\
         \midrule
         & Original Query & .444 & .184 & .245 & .092 & .437 & - & - \\
         & Original Query+RM3 & .447 & \textbf{.236}\ssa & .268 & .099 & .410 & - & 20 \\
         \midrule
         \multirow{2}{*}{Closed Source}
         & GPT4 & .467 & .201\ssa & .271 & .098 & .442 & 11.6 & 65  \\
         & GPT3.5 & \textbf{.486}\ssa & .213\ssa & .276 & \textbf{.115}\ssa & .440 & 1.8 & 23 \\
         \midrule
         \multirow{4}{*}{Open Source}
         & Qwen2-7B-Instruct & .476\ssa & .216\ssa & \textbf{.285}\ssa & .107 & \textbf{.465} & 2.6 & 45 \\
         & Phi3-medium-4k-Instruct & .453 & .218\ssa & .261 & .109 & .449 & 8.2 & 60 \\
         & Phi3-mini-4k-Instruct & .440 & .200 & .267 & .092 & .463 & 7.1 & 113 \\
         & Medical-Llama3-8B & .357\ssa & .222\ssa & .200 & .088 & .361 & 6.7 & 119 \\ 
         \midrule
         \multirow{1}{*}{SoA}
          & Neural Query Synthesis \cite{10.1145/3477495.3531853} & .473 & - & .276 & . & .434 & - & - 
    \end{tabular}%
    }
\end{table*} 
\begin{table*}[!t]
    \centering
    \caption{Results on the TREC22 benchmark using BM25 model. In boldface the best results overall. Blank cells indicate that the value can not be calculated or is not reported.}
    \label{tab:clinical_trial_trec22}
    \resizebox{\linewidth}{!}{%
    \begin{tabular}{cccccccp{7em}p{5em}}
         & Experiment & nDCG@10 & Bpref & P@10 & R@25 & MRR & Avg. processing time (seconds) & Average terms \\
         \midrule
         & Original Query & .409 & .177 & .276 & .101 & .561 & - & - \\
         & Original Query+RM3 & .457 & .248 & .354 & .127 & .504 & - & 20 \\
         \midrule
         \multirow{2}{*}{Closed Source}
         & GPT4 & .439 & .190\ssa & .304 & .107 & .571 & 9.2 & 47 \\
         & GPT35 & .496\ssa & .241\ssa & .342\ssa & \textbf{.140}\ssa & .655 & 2.1 & 15 \\
         \midrule
         \multirow{4}{*}{Open Source}
         & Qwen2-7B-Instruct & .509\ssa & .244\ssa & .370\ssa & .133\ssa & .627 & 1.7 & 29 \\
         & Phi-3-medium-4k-Instruct & \textbf{.532}\ssa & \textbf{.261}\ssa & \textbf{.400}\ssa & .138\ssa & \textbf{.698}\ssa & 8.1 & 63 \\
         & Phi3-mini-4k-Instruct & .420 & .229 & .296 & .111 & .589 &  6.7 & 127 \\
         & Medical-Llama3-8B & .357 & .214 & .272 & .108 & .428 & 7.3 & 111 \\
          \midrule
         \multirow{1}{*}{SoA}
          & frocchio run \cite{roberts2022overview} & .463 & - & .324 & . & .537 & - & -
    \end{tabular}%
    }
\end{table*} 

The obtained results suggest that the evaluated LLMs are effective for query generation for clinical trials retrieval, by consistently outperforming standard baselines and other approaches in the literature.
General-purpose LLMs like Qwen2-7B-Instruct, Phi-3-medium-4k-Instruct, and GPT3.5 have the best performance across both collections, with fast processing and moderate-term generation.
Notably, despite being significantly smaller than GPT models, these open-source LLMs achieve similar performance gains for the TREC21 collection and even greater improvements for TREC22.
More advanced (GPT4) and domain-specific models (Medical-Llama3-8B) underperform the other LLMs, exhibit longer response times, or generate more tokens.

The response times of the best performing LLMs are fast, ranging from approximately 1.7 to 8 seconds.
GPT3.5 and Qwen2-7B-Instruct have swift processing, with response times between 1.7 and 2.6 seconds, while Phi-3-medium-4k-Instruct takes 
longer (around 8 seconds).
These rapid response times are essential for the system's practical implementation, as they allow for near-instantaneous query expansion without introducing significant latency.
Regarding the number of generated terms, GPT3.5 generates only 15-23 terms on average, Qwen2-7B-Instruct produces a moderate number of terms (29-45),  while Phi-3-medium-4k-Instruct generates the 
highest number of terms (60-63).
In a real-world application, 
the limited number of terms generated allows medical experts to review a generated query by either accepting or modifying it
based on
their domain-specific knowledge.
This enables a synergy between AI-generated query expansions and human expertise in creating more accurate and contextually appropriate clinical trial queries.
\section{Discussion}
\label{sec:disc}

\begin{figure*}[!ht]
    \centering
    \includegraphics[width=\linewidth]{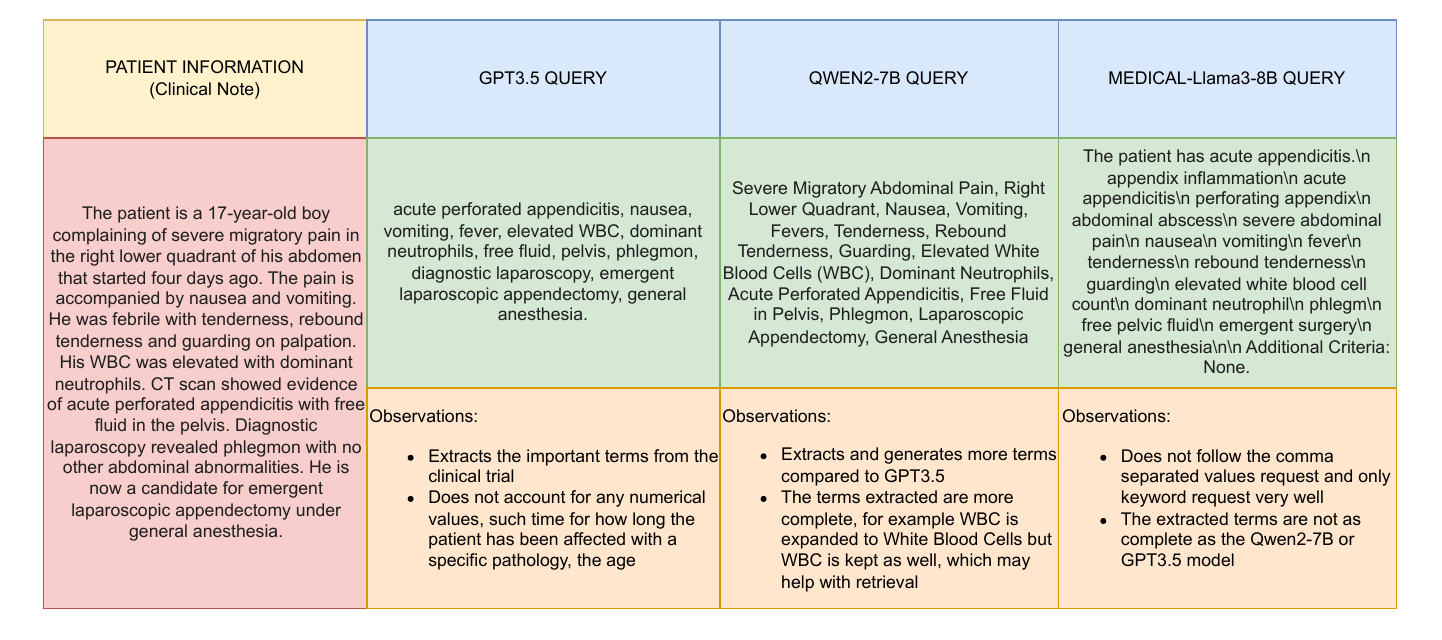}
    \caption{Qualitative example of LLM-generated queries and observations.}
    \label{fig:query_comparison}
\end{figure*}

The proposed system preserves the essential role of human expertise in identifying eligible clinical trials for patients. LLMs generate initial search queries that are then reviewed and refined by health experts, mitigating the already low risk of hallucinations in this application.
The risk of hallucination generation is minimized as the LLMs are instructed to extract specific medical terms directly from patient information rather than generating new content.
That is evidenced in our qualitative analysis, represented in Figure~\ref{fig:query_comparison}, where one can observe consistent reproduction of key clinical details across different models without introducing 
fake information.

Key advantages of the open-source LLMs evaluated in our study include their ability to operate effectively with minimal computational resources, to deliver performance comparable to or better than medical experts and larger models, and to address necessary privacy concerns.
The evaluated models, which rely on prompt instructions, perform better than 
the Medical-Llama3-8B, a domain-specific model but non fine-tuned for following prompt based instructions.
This also highlights the necessity of training domain-specific models to follow particular prompts instructions, e.g. for information extraction.

The clinical notes we used for evaluations are relatively short (7-10 sentences) and in English. However, the best performing models are multilingual and have significantly larger context windows.
For instance, Qwen2-7B-Instruct, trained in 29 languages including Chinese, exhibits high accuracy in extracting medical terminology from Chinese reports~\cite{info:doi/10.2196/54580}.
This multilingual capability, combined with its extensive context window of approximately 131k tokens, showcases the model's applicability in real-world settings across diverse linguistic healthcare contexts.
\section{Conclusions}
\label{sec:conc_fd}
The proposed system, which integrates LLMs into clinical trials retrieval, significantly enhances patient-trial matching effectiveness while maintaining privacy and preserving human supervision.
Our findings indicate that a proper usage of LLMs consistently outperforms standard baselines and existing approaches; in particular, small open-source LLMs like Qwen2-7B-Instruct and Phi3-medium-4k-Instruct achieve comparable or 
better performance than those achieved by both
larger closed-source models like GPT-3.5 and medical experts. These models balance performance and 
effectiveness, by offering fast processing times, manageable query lengths, and multilingual capabilities, making them highly suitable for practical implementation in real-world medical settings.

\section*{Acknowledgment}
This work was funded by the National Plan for NRRP Complementary Investments (PNC, established with the decree-law 6 May 2021, n. 59, converted by law n. 101 of 2021) in the call for the funding of research initiatives for technologies and innovative trajectories in the health and care sectors (Directorial Decree n. 931 of 06-06-2022) - project n. PNC0000003 - AdvaNced Technologies for Human-centrEd Medicine (project acronym: ANTHEM). This work reflects only the authors’ views and opinions, neither the Ministry for University and Research nor the European Commission can be considered responsible for them.
This work was partially supported by the European Union – Next Generation EU within the project NRPP M4C2, Investment 1.,3 DD. 341 - 15 march 2022 – FAIR – Future Artificial Intelligence Research – Spoke 4 - PE00000013 - D53C22002380006.

\bibliography{references}

\begin{thebibliography}{10}

\bibitem{DBLP:journals/csur/SarwarSCZAHVC23}
T.~Sarwar, S.~Seifollahi, J.~Chan, X.~Zhang, V.~Aksakalli, I.~L. Hudson, K.~Verspoor, and L.~Cavedon, ``The secondary use of electronic health records for data mining: Data characteristics and challenges,'' {\em {ACM} Comput. Surv.}, vol.~55, no.~2, pp.~33:1--33:40, 2023.

\bibitem{fleming1996surrogate}
T.~R. Fleming and D.~L. DeMets, ``Surrogate end points in clinical trials: are we being misled?,'' {\em Annals of internal medicine}, vol.~125, no.~7, pp.~605--613, 1996.

\bibitem{gul2010clinical}
R.~B. Gul and P.~A. Ali, ``Clinical trials: the challenge of recruitment and retention of participants,'' {\em Journal of clinical nursing}, vol.~19, no.~1-2, pp.~227--233, 2010.

\bibitem{jain2019conceptual}
N.~M. Jain, A.~Culley, T.~Knoop, C.~Micheel, T.~Osterman, and M.~Levy, ``Conceptual framework to support clinical trial optimization and end-to-end enrollment workflow,'' {\em JCO Clinical Cancer Informatics}, vol.~3, pp.~1--10, 2019.

\bibitem{DBLP:journals/kais/LandolsiHR23}
M.~Y. Landolsi, L.~Hlaoua, and L.~B. Romdhane, ``Information extraction from electronic medical documents: state of the art and future research directions,'' {\em Knowl. Inf. Syst.}, vol.~65, no.~2, pp.~463--516, 2023.

\bibitem{DBLP:journals/ijmi/HobensackSSBT23}
M.~Hobensack, J.~Song, D.~Scharp, K.~H. Bowles, and M.~Topaz, ``Machine learning applied to electronic health record data in home healthcare: {A} scoping review,'' {\em Int. J. Medical Informatics}, vol.~170, p.~104978, 2023.

\bibitem{DBLP:conf/ecir/PeikosAPV23}
G.~Peikos, D.~Alexander, G.~Pasi, and A.~P. de~Vries, ``Investigating the impact of query representation on medical information retrieval,'' in {\em Advances in Information Retrieval - 45th European Conference on Information Retrieval, {ECIR} 2023, Ireland, Proceedings, Part {II}}, vol.~13981 of {\em Lecture Notes in Computer Science}, pp.~512--521, Springer, 2023.

\bibitem{DBLP:journals/jbi/ParkFTZICFSGSW24}
J.~Park, Y.~Fang, C.~N. Ta, G.~Zhang, B.~Idnay, F.~Chen, D.~Feng, R.~Shyu, E.~R. Gordon, M.~E. Spotnitz, and C.~Weng, ``Criteria2query 3.0: Leveraging generative large language models for clinical trial eligibility query generation,'' {\em J. Biomed. Informatics}, vol.~154, p.~104649, 2024.

\bibitem{peikos2023utilizing}
G.~Peikos, S.~Symeonidis, P.~Kasela, and G.~Pasi, ``Utilizing chatgpt to enhance clinical trial enrollment,'' {\em arXiv preprint arXiv:2306.02077}, 2023.

\bibitem{zuhair2024exploring}
V.~Zuhair, A.~Babar, R.~Ali, M.~O. Oduoye, Z.~Noor, K.~Chris, I.~I. Okon, and L.~U. Rehman, ``Exploring the impact of artificial intelligence on global health and enhancing healthcare in developing nations,'' {\em Journal of Primary Care \& Community Health}, vol.~15, p.~21501319241245847, 2024.

\bibitem{patel2023chatgpt}
S.~B. Patel and K.~Lam, ``Chatgpt: the future of discharge summaries?,'' {\em The Lancet Digital Health}, vol.~5, no.~3, pp.~e107--e108, 2023.

\bibitem{DBLP:journals/jms/CascellaMBB23}
M.~Cascella, J.~Montomoli, V.~Bellini, and E.~Bignami, ``Evaluating the feasibility of chatgpt in healthcare: An analysis of multiple clinical and research scenarios,'' {\em J. Medical Syst.}, vol.~47, no.~1, p.~33, 2023.

\bibitem{sezgin2022operationalizing}
E.~Sezgin, J.~Sirrianni, and S.~L. Linwood, ``Operationalizing and implementing pretrained, large artificial intelligence linguistic models in the us health care system: outlook of generative pretrained transformer 3 (gpt-3) as a service model,'' {\em JMIR medical informatics}, vol.~10, no.~2, p.~e32875, 2022.

\bibitem{DBLP:journals/corr/abs-2303-11032}
Z.~Liu, X.~Yu, L.~Zhang, Z.~Wu, C.~Cao, H.~Dai, L.~Zhao, W.~Liu, D.~Shen, Q.~Li, T.~Liu, D.~Zhu, and X.~Li, ``Deid-gpt: Zero-shot medical text de-identification by {GPT-4},'' {\em CoRR}, vol.~abs/2303.11032, 2023.

\bibitem{harrer2023attention}
S.~Harrer, ``Attention is not all you need: the complicated case of ethically using large language models in healthcare and medicine,'' {\em Ebiomedicine}, vol.~90, 2023.

\bibitem{yang2023large}
R.~Yang, T.~F. Tan, W.~Lu, A.~J. Thirunavukarasu, D.~S.~W. Ting, and N.~Liu, ``Large language models in health care: Development, applications, and challenges,'' {\em Health Care Science}, vol.~2, no.~4, pp.~255--263, 2023.

\bibitem{DBLP:journals/jamia/WuRDDJSSWWXZX20}
S.~Wu, K.~Roberts, S.~Datta, J.~Du, Z.~Ji, Y.~Si, S.~Soni, Q.~Wang, Q.~Wei, Y.~Xiang, B.~Zhao, and H.~Xu, ``Deep learning in clinical natural language processing: a methodical review,'' {\em J. Am. Medical Informatics Assoc.}, vol.~27, no.~3, pp.~457--470, 2020.

\bibitem{DBLP:journals/jbi/KreimeyerFPAHJF17}
K.~Kreimeyer, M.~Foster, A.~Pandey, N.~Arya, G.~Halford, S.~F. Jones, R.~Forshee, M.~Walderhaug, and T.~Botsis, ``Natural language processing systems for capturing and standardizing unstructured clinical information: {A} systematic review,'' {\em J. Biomed. Informatics}, vol.~73, pp.~14--29, 2017.

\bibitem{DBLP:journals/jbi/WangWRMSALZMSL18}
Y.~Wang, L.~Wang, M.~Rastegar{-}Mojarad, S.~Moon, F.~Shen, N.~Afzal, S.~Liu, Y.~Zeng, S.~Mehrabi, S.~Sohn, and H.~Liu, ``Clinical information extraction applications: {A} literature review,'' {\em J. Biomed. Informatics}, vol.~77, pp.~34--49, 2018.

\bibitem{alsentzer-etal-2019-publicly}
E.~Alsentzer, J.~Murphy, W.~Boag, W.-H. Weng, D.~Jindi, T.~Naumann, and M.~McDermott, ``Publicly available clinical {BERT} embeddings,'' in {\em Proceedings of the 2nd Clinical Natural Language Processing Workshop}, (Minneapolis, Minnesota, USA), pp.~72--78, Association for Computational Linguistics, June 2019.

\bibitem{beltagy2019scibert}
I.~Beltagy, K.~Lo, and A.~Cohan, ``{S}ci{BERT}: A pretrained language model for scientific text,'' in {\em Proceedings of the 2019 Conference on Empirical Methods in Natural Language Processing and the 9th International Joint Conference on Natural Language Processing (EMNLP-IJCNLP)}, (Hong Kong, China), pp.~3615--3620, Association for Computational Linguistics, Nov. 2019.

\bibitem{hahn2020medical}
U.~Hahn and M.~Oleynik, ``Medical information extraction in the age of deep learning,'' {\em Yearbook of medical informatics}, vol.~29, no.~01, pp.~208--220, 2020.

\bibitem{hu2023zero}
Y.~Hu, I.~Ameer, X.~Zuo, X.~Peng, Y.~Zhou, Z.~Li, Y.~Li, J.~Li, X.~Jiang, and H.~Xu, ``Zero-shot clinical entity recognition using chatgpt,'' {\em arXiv preprint arXiv:2303.16416}, 2023.

\bibitem{DBLP:conf/emnlp/AgrawalHLKS22}
M.~Agrawal, S.~Hegselmann, H.~Lang, Y.~Kim, and D.~A. Sontag, ``Large language models are few-shot clinical information extractors,'' in {\em {EMNLP}}, pp.~1998--2022, Association for Computational Linguistics, 2022.

\bibitem{huang2024critical}
J.~Huang, D.~M. Yang, R.~Rong, K.~Nezafati, C.~Treager, Z.~Chi, S.~Wang, X.~Cheng, Y.~Guo, L.~J. Klesse, {\em et~al.}, ``A critical assessment of using chatgpt for extracting structured data from clinical notes,'' {\em npj Digital Medicine}, vol.~7, no.~1, p.~106, 2024.

\bibitem{choi2023developing}
H.~S. Choi, J.~Y. Song, K.~H. Shin, J.~H. Chang, and B.-S. Jang, ``Developing prompts from large language model for extracting clinical information from pathology and ultrasound reports in breast cancer,'' {\em Radiation Oncology Journal}, vol.~41, no.~3, p.~209, 2023.

\bibitem{li2024scoping}
L.~Li, J.~Zhou, Z.~Gao, W.~Hua, L.~Fan, H.~Yu, L.~Hagen, Y.~Zhang, T.~L. Assimes, L.~Hemphill, {\em et~al.}, ``A scoping review of using large language models (llms) to investigate electronic health records (ehrs),'' {\em arXiv preprint arXiv:2405.03066}, 2024.

\bibitem{soboroff2021overview}
I.~Soboroff, ``Overview of trec 2021,'' in {\em 30th Text REtrieval Conference. Gaithersburg, Maryland}, 2021.

\bibitem{roberts2022overview}
K.~Roberts, D.~Demner-Fushman, E.~M. Voorhees, S.~Bedrick, and W.~R. Hersh, ``Overview of the trec 2022 clinical trials track.,'' in {\em TREC}, 2022.

\bibitem{DBLP:conf/sigir/KoopmanZ16}
B.~Koopman and G.~Zuccon, ``A test collection for matching patients to clinical trials,'' in {\em Proceedings of the 39th International {ACM} {SIGIR} conference on Research and Development in Information Retrieval, {SIGIR} 2016}, pp.~669--672, {ACM}, 2016.

\bibitem{DBLP:conf/sigir/AgostiN019}
M.~Agosti, G.~M.~D. Nunzio, and S.~Marchesin, ``An analysis of query reformulation techniques for precision medicine,'' in {\em Proceedings of the 42nd International {ACM} {SIGIR} Conference on Research and Development in Information Retrieval, {SIGIR} 2019, Paris, France, July 21-25, 2019} (B.~Piwowarski, M.~Chevalier, {\'{E}}.~Gaussier, Y.~Maarek, J.~Nie, and F.~Scholer, eds.), pp.~973--976, {ACM}, 2019.

\bibitem{DBLP:journals/jbi/RybinskiXK20}
M.~Rybinski, J.~Xu, and S.~Karimi, ``Clinical trial search: Using biomedical language understanding models for re-ranking,'' {\em J. Biomed. Informatics}, vol.~109, p.~103530, 2020.

\bibitem{10.1145/3477495.3531853}
R.~Pradeep, Y.~Li, Y.~Wang, and J.~Lin, ``Neural query synthesis and domain-specific ranking templates for multi-stage clinical trial matching,'' in {\em Proceedings of the 45th International ACM SIGIR Conference on Research and Development in Information Retrieval}, SIGIR '22, (New York, NY, USA), p.~2325–2330, Association for Computing Machinery, 2022.

\bibitem{DBLP:conf/trec/RobertsonWJHG94}
S.~E. Robertson, S.~Walker, S.~Jones, M.~Hancock{-}Beaulieu, and M.~Gatford, ``Okapi at {TREC-3},'' in {\em Proceedings of The Third Text REtrieval Conference, {TREC} 1994, Gaithersburg, Maryland, USA, November 2-4, 1994} (D.~K. Harman, ed.), vol.~500-225 of {\em {NIST} Special Publication}, pp.~109--126, National Institute of Standards and Technology {(NIST)}, 1994.

\bibitem{DBLP:conf/trec/JaleelACDLLSW04}
N.~A. Jaleel, J.~Allan, W.~B. Croft, F.~Diaz, L.~S. Larkey, X.~Li, M.~D. Smucker, and C.~Wade, ``Umass at {TREC} 2004: Novelty and {HARD},'' in {\em Proceedings of the Thirteenth Text REtrieval Conference, {TREC} 2004, Gaithersburg, Maryland, USA, November 16-19, 2004} (E.~M. Voorhees and L.~P. Buckland, eds.), vol.~500-261 of {\em {NIST} Special Publication}, National Institute of Standards and Technology {(NIST)}, 2004.

\bibitem{grootendorst2020keybert}
M.~Grootendorst, ``Keybert: Minimal keyword extraction with bert.,'' 2020.

\bibitem{bodenreider2004unified}
O.~Bodenreider, ``The unified medical language system (umls): integrating biomedical terminology,'' {\em Nucleic acids research}, vol.~32, no.~suppl\_1, pp.~D267--D270, 2004.

\bibitem{peikos2022unimib}
G.~Peikos, O.~Espitia, and G.~Pasi, ``Unimib at trec 2021 clinical trials track,'' {\em arXiv preprint arXiv:2207.13514}, 2022.

\bibitem{kusa2021dossier}
W.~Kusa and Y.~Ghafourian, ``Dossier at trec 2021 clinical trials track.,'' in {\em TREC}, 2021.

\bibitem{reverberi2022experimental}
C.~Reverberi, T.~Rigon, A.~Solari, C.~Hassan, P.~Cherubini, and A.~Cherubini, ``Experimental evidence of effective human--ai collaboration in medical decision-making,'' {\em Scientific reports}, vol.~12, no.~1, p.~14952, 2022.

\bibitem{shah2020secondary}
S.~M. Shah and R.~A. Khan, ``Secondary use of electronic health record: Opportunities and challenges,'' {\em IEEE access}, vol.~8, pp.~136947--136965, 2020.

\bibitem{yang2024qwen2technicalreport}
A.~Y. et~al., ``Qwen2 technical report,'' 2024.

\bibitem{abdin2024phi3technicalreport}
M.~A. et~al., ``Phi-3 technical report: A highly capable language model locally on your phone,'' 2024.

\bibitem{pyterrier2020ictir}
C.~Macdonald and N.~Tonellotto, ``Declarative experimentation ininformation retrieval using pyterrier,'' in {\em Proceedings of ICTIR 2020}, 2020.

\bibitem{DBLP:books/daglib/0021593}
C.~D. Manning, P.~Raghavan, and H.~Sch{\"{u}}tze, {\em Introduction to information retrieval}.
\newblock Cambridge University Press, 2008.

\bibitem{DBLP:journals/ftir/MitraC18}
B.~Mitra and N.~Craswell, ``An introduction to neural information retrieval,'' {\em Found. Trends Inf. Retr.}, vol.~13, no.~1, pp.~1--126, 2018.

\bibitem{DBLP:conf/sigir/Sakai07}
T.~Sakai, ``Alternatives to bpref,'' in {\em {SIGIR} 2007: Proceedings of the 30th Annual International {ACM} {SIGIR} Conference on Research and Development in Information Retrieval, Amsterdam, The Netherlands, July 23-27, 2007} (W.~Kraaij, A.~P. de~Vries, C.~L.~A. Clarke, N.~Fuhr, and N.~Kando, eds.), pp.~71--78, {ACM}, 2007.

\bibitem{carterette2012multiple}
B.~A. Carterette, ``Multiple testing in statistical analysis of systems-based information retrieval experiments,'' {\em ACM Transactions on Information Systems (TOIS)}, vol.~30, no.~1, pp.~1--34, 2012.

\bibitem{info:doi/10.2196/54580}
L.~Wang, Y.~Ma, W.~Bi, H.~Lv, and Y.~Li, ``An entity extraction pipeline for medical text records using large language models: Analytical study,'' {\em J Med Internet Res}, vol.~26, p.~e54580, Mar 2024.

\end{thebibliography}
\bibliographystyle{ieeetr}
\end{document}